\documentclass[epsfig,12pt]{article}
\usepackage{epsfig}
\usepackage{graphicx}
\usepackage{array}
\usepackage{caption}
\usepackage{subcaption}
\usepackage{slashed}
\usepackage{amsmath}

\newcommand{\beq}{\begin{equation}}   
\newcommand{\eeq}{\end{equation}}
\newcommand{\beqn}{\begin{eqnarray}}   
\newcommand{\eeqn}{\end{eqnarray}}

\newcommand{\bea}{\begin{eqnarray}}
\newcommand{\eea}{\end{eqnarray}}
\newcommand{\be}{\begin{equation}}
\newcommand{\ee}{\end{equation}}
\newcommand{\bead}{\begin{aligned}}
\newcommand{\eead}{\end{aligned}}

\newcommand{\gsim}{\lower.7ex\hbox{$
\;\stackrel{\textstyle>}{\sim}\;$}}
\newcommand{\lsim}{\lower.7ex\hbox{$
\;\stackrel{\textstyle<}{\sim}\;$}}
\begin{document}

\begin{titlepage}

\begin{flushright}
FTPI-MINN-14/2, UMN-TH-3323/14
\end{flushright}

\vspace{0.7cm}

\begin{center}
{  \large \bf  More on the Abrikosov Strings with Non-Abelian \\[2mm]Moduli}
\end{center}
\vspace{0.6cm}

\begin{center}
 {\large 
 M. Shifman,$^a$ Gianni Tallarita,$^b$ and Alexei Yung$^{a,c}$}
\end {center}

\vspace{3mm}
 
\begin{center}

$^a${\em William I. Fine Theoretical Physics Institute, University of Minnesota,
Minneapolis, MN 55455, USA}\\[1mm]
$^b${\em Centro de Estudios Cient\'{i}ficos (CECs), Casilla 1469, Valdivia, Chile}
\\[1mm]
$^{c}${\em Petersburg Nuclear Physics Institute, Gatchina, St. Petersburg
188300, Russia}

\end {center}

\vspace{2cm}

\begin{center}
{\large\bf Abstract}
\end{center}

We continue explorations of deformed Abrikosov-Nielsen-Olesen (ANO) strings, with non-Abelian moduli
on the world sheet. In a simple model with an extra field we find classically stable ANO and non-Abelian strings.
The tension of the latter is a few percent lower than the tension of the ANO string. Then we calculate
the interpolating field configuration.
Once the kink mass $M_k$ and the difference of tensions $\Delta T$ are found we  calculate the decay rate of the ANO string with a higher tension (``false vacuum") into the non-Abelian string with the lower tension (``genuine vacuum")
through the ``bubble" formation in the quasiclassical approximation.

\hspace{0.3cm}

\end{titlepage}


\section{Introduction}
 \label{intro}

This paper is a continuation of the studies of generalized Abrikosov-Nielsen-Olesen
 strings \cite{ANO} 
 with non-Abelian moduli on the world sheet
\cite{S,Sh,MSY,Monin:2013mxa}. Our main task is two-fold: (a) search for new stable and (nearly) degenerate string
solutions with and without non-Abelian moduli, respectively, (see \cite{Monin:2013mxa}); and
(b) calculation of the mass of the kink interpolating between these two (nearly) degenerate strings.

Once the kink mass $M_k$ and the difference of tensions $\Delta T$ are found we can calculate the decay rate of the string with a higher tension (``false vacuum") into the string with the lower tension (``genuine vacuum")
through the ``bubble" formation (see e.g. \cite{SA}). 
Both points (a) and (b) are studied in the framework of the model \cite{S}.
Quasiclassical consideration requires the bulk theories to be weakly coupled, a condition which will be met by an appropriate choice of the coupling constants. In addition, 
calculation
of the string decay rate requires $M_k^2/\Delta T$ to be large.

In the historical perspective, topologically stable non-Abelian strings were first found in \cite{AHDT}. Our starting point is the model \cite{S}  described by the Lagrangian
\beq
{\cal L} = {\cal L}_{0} +{\cal L}_\chi
\label{odin}
\eeq
where
 \beqn
{\cal L}_{0} &=& -\frac{1}{4e^2}F_{\mu\nu}^2 + \left| {\mathcal D}^\mu\phi\right|^2
  -V(\phi )\, ,
  \nonumber\\[2mm]
  {\mathcal D}_\mu\phi &=& (\partial_\mu -iA_\mu )\phi\,,
\nonumber\\[2mm]
  V &=& \lambda \left(|\phi |^2 -v^2
\right)^2\,,
  \label{tpi16}
\eeqn
and
\beqn
{\cal L}_\chi &=& \partial_\mu \chi^i \, \partial^\mu \chi^i  - U(\chi, \phi)\,,
\label{14}\\[2mm]
U &=&  \gamma\left[\left(-\mu^2 +|\phi |^2
\right)\chi^i \chi^i + \beta \left( \chi^i \chi^i\right)^2\right],
\label{15p}
\eeqn
with $i=1,2,3$. We will assume that $\lambda, \beta ,\gamma  > 0$ and $v^2 > \mu^2$. This model has the U(1) gauge symmetry, while the $\chi$ sector is O(3) symmetric. The $\chi$ fields are assumed to be real.
This model is similar to the Witten's model for superconducting strings \cite{Wsupcond}.

The potential $V(\phi)$ ensures the Higgsing of the $U(1)$ photon. The complex scalar field $\phi$ develops a 
nonvanishing vacuum expectation value (VEV), 
$$|\phi|=v\,.$$
 As a result of the Higgs mechanism the phase of the complex field is eaten up and becomes photon's longitudinal component. The photon   mass is
\beq
m^2_A=2e^2v^2\,.
\label{h6} 
\eeq
The physical Higgs excitation obviously has the  mass
\beq
m^2_{\phi}=4\lambda v^2\, .
\label{h7} 
\eeq
As can be seen from (\ref{15p}), the triplet field $\chi^i$ does {\em not} condense in the vacuum. The mass 
of the $\chi$ quantum is
\beq
m^2_{\chi}=\gamma\left(v^2-\mu^2\right).
\label{h8}
\eeq

For what follows it is convenient to introduce three auxiliary dimensionless parameters:
\beq
a=\frac{m^2_A}{m^2_{\phi}}\equiv\frac{e^2}{2\lambda}\,,\quad b=\frac{m^2_{\chi}}{m^2_\phi} \equiv \frac{\gamma}{4\lambda}\,\frac{c-1}{c}\,,\quad c=\frac{v^2}{\mu^2}\,.
\eeq

As was discussed in \cite{MSY},  a constraint  on the parameters of the Lagrangian exists from the requirement of vacuum stability, namely,
\beq
\beta \geq \beta_*\equiv\frac{b}{c(c-1)}\,.
\eeq

\vspace{0.5cm}
\newcolumntype{C}{ >{\centering\arraybackslash} p{3cm} }
\newcolumntype{D}{ >{\centering\arraybackslash} p{3cm} }
\begin{table}[h!]
\centering
\renewcommand\arraystretch{3.5}
\begin{tabular}{|C|D|}
\hline
$\beta$ & $\displaystyle\frac{\tilde\lambda}{\gamma}$ \\\hline
$a$ & $\displaystyle\frac{m_A^2}{m_{\phi}^2}$ \\\hline
$b$ & $\displaystyle\frac{m_{\chi}^2}{m_{\phi}^2}$ \\\hline
$\displaystyle\frac{v^2}{\mu^2}\equiv c$ & $\displaystyle \left( 1-\frac{4\lambda}{\gamma}\frac{m_{\chi}^2}{m_{\phi}^2}\right)^{-1}$ \\\hline
\end{tabular}
\caption{\small Parameters in (\ref{tpi16}), (\ref{15p}) in terms of the particle masses and the coefficients in front of 
the quartic terms $\phi^4$, $\chi^4$, and $\phi^2\chi^2$  ($\lambda$, $\tilde\lambda$, and $\gamma$, respectively).}
\label{t1}
\end{table}

Table \ref{t1} shows how the parameters in (\ref{tpi16}), (\ref{15p}) and $a,b,c$ are related  to the masses  
of the particles involved and to the 
 coefficients in front of the quartic terms $\phi^4$, $\chi^4$ and $\phi^2\chi^2$. 
 
 Organization of the paper is as follows.
In section 2 we continue the studies initiated in \cite{S,Sh,MSY,Monin:2013mxa} and, after proving the stability of the abelian string solution which is higher in tension than its non-abelian counterpart, find a ``kink"-like interpolating solution between the two. In this section we also calculate the decay rate of one string into the other. In section 3 we  analyze of the low-energy effective theory on the string world sheet. 

   \section{More on (nearly) degenerate strings}
   \label{mon} 
 
 In \cite{Monin:2013mxa} solitonic solutions of the model described by Lagrangian (\ref{odin}) were found in which the ANO string acquires extra orientational moduli. These strings were shown to be approximately degenerate in tension to their Abelian counterparts, the ANO strings.
The value of $\chi$ detected in the core of the string solution, although nonvanishing, was rather small.
In this section we want to extend these solutions. Scanning the parameter space we find  numerical solutions where $\chi$ is not small in the core. We will show that even though a larger value of $\chi$ in the core takes one further away from the tension degeneracy, nonetheless one can, under suitable approximations, find interpolating solutions between the two strings (i.e. kinks). 
Note that the ANO string (i.e. the $\chi\equiv 0$ string) always exists in the model at hand. We will need only to show its classical stability. The non-Abelian string solution ($\chi\neq 0$ in the  string core) to be presented below will have a lower tension than the ANO string.

\subsection{Non-Abelian solution}
\label{nas}
 
 First, let us present the non-Abelian string solutions. Using the ansatz
 \beqn
A_i &=& -\epsilon_{ij}\frac{x_j}{r^2}(n_e-f(r))\,,
\nonumber \\[2mm]
\phi &=& v\varphi(r)e^{in_e\theta}\,,
\nonumber \\[2mm]
\chi^i &=& \frac{\mu}{\sqrt{2\beta}}\chi(r)(0,0,1)
\eeqn
 and introducing the dimensionless parameter
  $$\rho = m_\phi r$$
   one finds the energy minimization equations (here the prime denotes differentiation with respect to $\rho$)
 \beqn
&&(\rho \varphi')'-\frac{\rho}{2}\varphi(\varphi^2-1)-\frac{1}{\rho}f^2\varphi-\frac{b\rho}{2\beta(c-1)}\chi^2\varphi=0
\nonumber \\[2mm]
&&(\rho\chi')'-\frac{b\rho}{(c-1)}\chi\left[\chi^2+c\varphi^2-1\right]=0
\nonumber \\[2mm]
&&\left(f'/\rho\right)'-\frac{a\varphi^2}{\rho}f=0
\eeqn
which we need to solve under  the following boundary conditions:
\beqn
&&\chi'(0) =0,\qquad \chi (\infty) =0\,,
\nonumber \\[2mm]
&&\varphi(0)=0, \qquad \varphi(\infty) = 1\,,
\nonumber \\[2mm]
&&f(0)=n_e\,, \qquad f'(\infty) = 0\,.
\eeqn
Solutions are presented in Figure \ref{fig222}. They correspond to winding number
$n_e=1$.
Note that the value of $\chi$ in the core of the string is not small for these solutions.  The value of the string tension for this solution is
\beq
\frac{T_{\rm NA}}{2\pi v^2} \approx 0.87.
\eeq
Thus, the difference in string tensions to be used below is
\beq
\frac{\Delta T}{2\pi v^2} \equiv \frac{T_{ANO} - T_{NA}}{2\pi v^2} \approx  0.13.
\eeq

 \subsection{Classical stability of the ANO string}
 \label{cso}
 
 As was noted in \cite{Monin:2013mxa},
 the tension of the ANO string $T_{\rm ANO}$ in the model at hand is higher 
 than $T_{\rm NA}$.
 Whether or not both can coexist at the classical level -- i.e. whether or not the ANO string is quasistable -- depends on its classical stability. We need to prove that there are no negative modes in the background of the ANO solution.
 Here we will present such a proof. \newline
 
 The energy functional for $\chi$ in the quadratic approximation is
 \be\label{energy1}
 \mathcal{E}_\chi = \frac{\mu^2}{2\beta}L\int dx dy \left\{\chi \left[-\Delta+\gamma\mu^2\left(-1+\frac{v^2}{\mu^2}\varphi_0^2\right)\right]\chi \right\}
 \ee
 where $L$ is the string length (tending to infinity). From this, introducing 
 \be
 \psi(\rho)=\sqrt{\rho}\chi
 \ee
  one can derive a one-dimensional Schr\"odinger equation for the $\chi$ modes
 \be\label{schro}
 -\psi''+\left(b\frac{c\varphi_0^2-1}{c-1}-\frac{1}{4\rho^2}\right)\psi=\epsilon\psi, \quad \epsilon = \frac{E}{m_\phi^2}.
 \ee
Using the solution for $\varphi_0 = \varphi_{ANO}$ at $b=0.1$, $c=1.07$ we find the lowest-lying mode is at $\epsilon\approx 0.1042$. The positivity of $\epsilon$ demonstrates the classical stability of the $\chi= 0$ solution.

 \subsection{Interpolating kink solution}
 \label{iks}
 
  If the degeneracy of the string tensions was exact, there would exist a static kink solution interpolating between the ANO string (with no orientational moduli on the world sheet) and the non-Abelian string. However,  their  tensions are not exactly
   degenerate.  Hence no static kink solution exists in this case. When looking for  an interpolating solution one 
   needs to take into account the fact that the kink will move in the direction of the ANO string. This will lower the 
   energy of the interpolating field configuration. Thus, one
   is forced to include the time dependence of the fields. 
   
   Given that (to the degree of accuracy of our solutions) the non-degeneracy in tensions is relatively small,
   one can make reasonable assumptions which will simplify the problem greatly. First, the variation of the gauge and 
   $\phi$ fields between the solutions with $\chi \equiv 0$ and those with $\chi\neq 0$ can be neglected. Secondly the force that the kink will experience will be small and, hence, its acceleration can be taken as a small parameter. These two assumptions -- no dependence of the $\phi$ and $f$ fields on the string world sheet coordinate $z$ and time $t$ and small acceleration of the $\chi$ field -- turn out to be sufficient in order to find an approximate kink solution interpolating between both strings. \newline
 
 Let us introduce a time dependence for the $\chi$ field in the following form:
 \be\label{acc}
\frac{1}{m_\phi^2}\frac{d^2}{dt^2}\chi\left(\rho, z-\frac{1}{2} \tilde{a}t^2\right)\approx -\hat{a}\partial_{\tilde{z}}\chi
\ee
where $$\tilde{z}=m_\phi \left(z-\frac{1}{2} \tilde{a}t^2\right)$$ and $\hat{a}=\tilde{a}/m_\phi$. In deriving (\ref{acc}) we made use of the small acceleration approximation. Then, with the assumptions described above, we can solve the two-dimensional problem
\beqn
&& (\rho \varphi')'-\frac{\rho}{2}\varphi(\varphi^2-1)-\frac{1}{\rho}f^2\varphi-\frac{b\rho}{2\beta(c-1)}\chi^2\varphi=0\,,
\nonumber\\[2mm]
&&(\rho\chi')'+\rho\;\partial^2_{\tilde{z}} \chi-\hat{a}\;\partial_{\tilde{z}}\chi-\frac{b\rho}{(c-1)}\chi\left[\chi^2+c\varphi^2-1\right]=0\,,
\nonumber\\[2mm]
&&\left(f'/\rho\right)'-\frac{a\varphi^2}{\rho}f=0\,,
\eeqn
with the boundary conditions
\beqn
&&\chi'(0,\tilde{z}) =0,\qquad \chi (\infty,\tilde{z}) =0
\nonumber\\[2mm]
&&\chi(\rho,-\infty) =0,\qquad \chi'(\rho,\infty) =0
\nonumber\\[2mm]
&&\varphi(0)=0, \qquad \varphi(\infty) = 1\,,
\nonumber\\[2mm]
&& f(0)=n_e\,, \qquad f'(\infty) = 0\,.
\eeqn
The interpolating solution is  presented in Figure \ref{fig2}, where $n_e = 1$.
 
 \subsection{Kink mass and the string decay rate} 
 \label{kma}
 
 First, we will determine the kink mass.
 
 From the energy plot presented in Figure \ref{fig:sub1} we determine the kink to $\chi$ mass ratio by looking at the excess energy (the ``bump" energy) with respect to a smooth unstable decay towards $\chi=0$. We have
\be
\frac{\delta T}{2\pi v^2} \tilde{z} = M_k \times \frac{\sqrt{\lambda}}{\pi v} \approx 0.1,
\ee
where $\delta T$ represents the excess energy only, but since
\be
v = m_\chi\sqrt{\frac{c}{\gamma(c-1)}}
\ee
then
\be
\frac{M_k}{m_{\chi}}\approx0.1\times\frac{\pi}{2\lambda}\sqrt{\frac{1}{b}}=\frac{3.8}{\lambda}\gg 1\,,
\ee
where the last equality holds because the parameter $\lambda$ can be chosen at will  and must be small for the classical regime to be reliable.  As was expected, $M_k \sim m_\chi/\lambda$.
\newline

From this plot one can also determine the decay rate of the Abelian string into a non-Abelian one.
 This will happen by nucleation of a non-Abelian string joined by a pair of ``kinks" to an Abelian solution,
 which is higher in energy. Once this solution is formed the ``bubble" corresponding to the non-Abelian 
string will expand classically. This configuration will have an effective action
 \be
 S_{\rm eff} = 2\pi (M_k) L-\pi L^2\Delta T
 \ee
 where the first term represents the tension of the wall (in this case the energy of the two kinks) and the second term the gain in energy in passing to the lower energy solution. The parameter $L$ is the ``radius" of the bubble (this involves the time coordinate). Minimization of this energy with respect to $L$ gives the critical action
 \be
 S^*_{\rm eff} = \frac{\pi M_k^2}{\Delta T}.
 \ee
 The decay rate will therefore be \cite{SA}
 \be
 \Gamma \approx \exp\left({-S^*_{\rm eff}}\right) = \exp\left({-\frac{\pi M_k^2}{\Delta T}}\right).
 \ee
 From our plot we find
  \beq
 \Gamma \approx \exp\left({-\frac{\pi M_k^2}{\Delta T}}\right) = \exp\left(-2\lambda b\frac{(M_k/m_\chi)^2}{\Delta T/2\pi v^2}\right) \approx  \exp\left(-\frac{22.5}{\lambda}\right).
 \eeq

 \section{Low-energy theory on the string world sheet}
 \label{let}

The rotational moduli on the generalized ANO string world-sheet were introduced  in \cite{MSY}. 
We will follow the same line of reasoning.

\begin{figure}
\centering
\begin{subfigure}{.5\textwidth}
  \centering
  \includegraphics[width=0.9\linewidth]{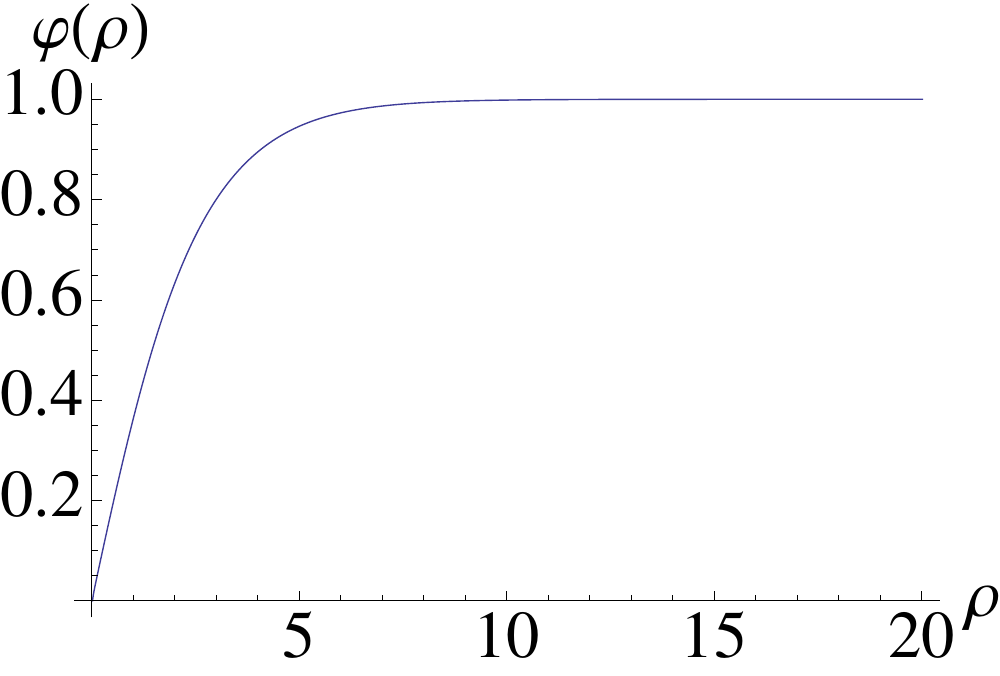}
  \caption{$\varphi$ profile}
  \label{fig:sub1}
\end{subfigure}%
\begin{subfigure}{.5\textwidth}
  \centering
  \includegraphics[width=.9\linewidth]{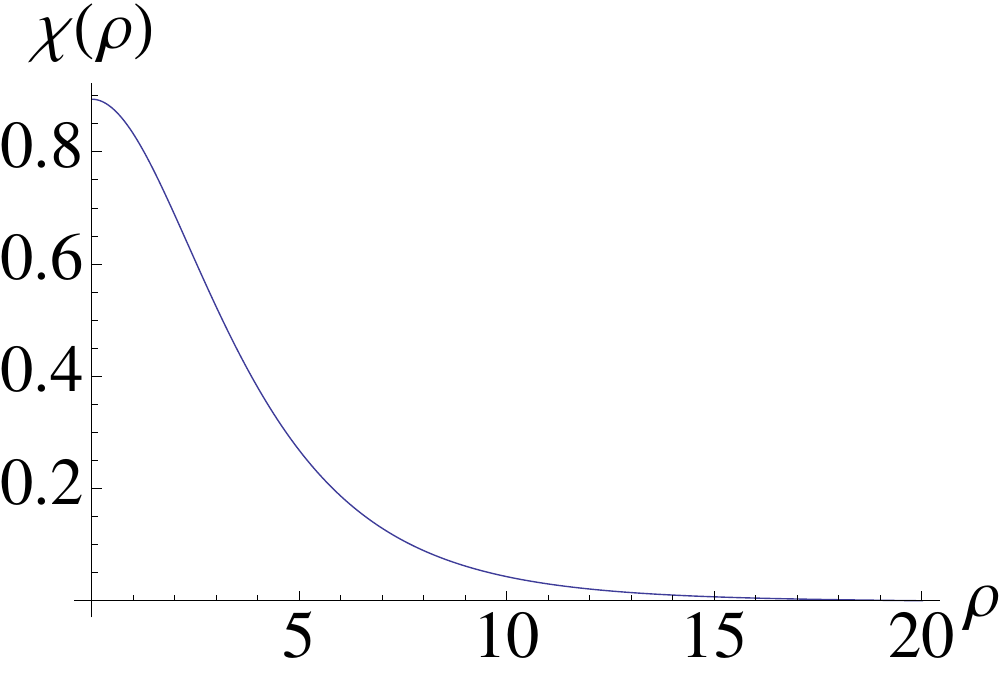}
  \caption{$\chi$ profile}
  \label{fig:sub2222}
\end{subfigure}
\label{fig:test}
\centering
\begin{subfigure}{.5\textwidth}
  \centering
  \includegraphics[width=\linewidth]{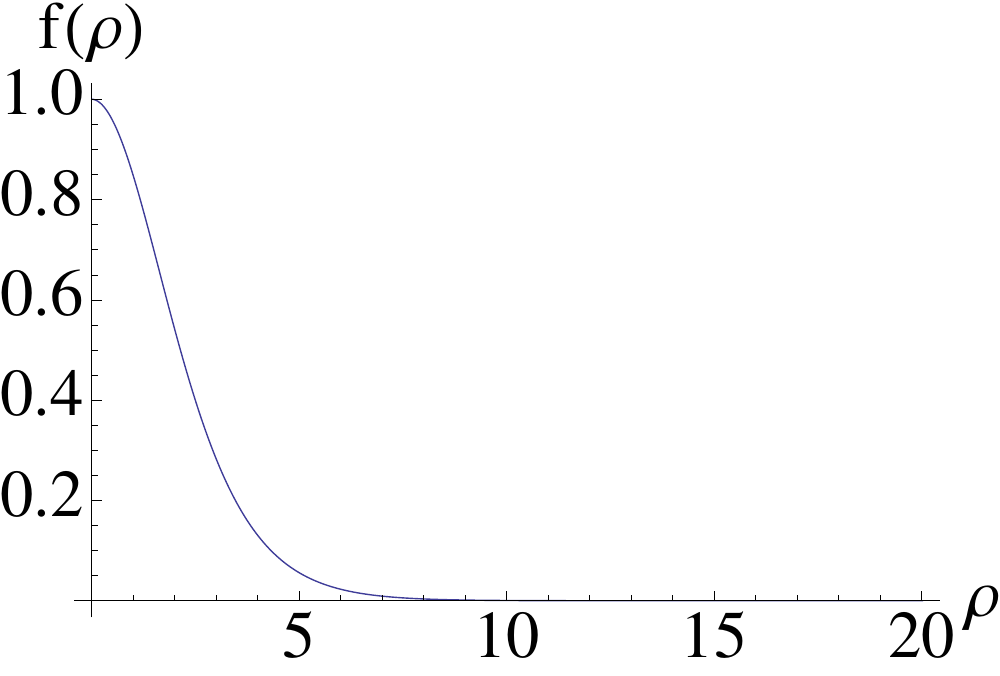}
  \caption{$f$ profile}
  \label{fig:subaaa1}
\end{subfigure}%
\caption{Field profiles for the non-abelian string solution. The plots correspond to the numerical choices of $a =1, b=0.1$, $c=1.07$, $\beta = 1.1 b/(c(c-1))$, $n_e=1$.}
\label{fig222}
\end{figure}

\begin{figure}
\centering
\begin{subfigure}{.5\textwidth}
  \centering
  \includegraphics[width=0.9\linewidth]{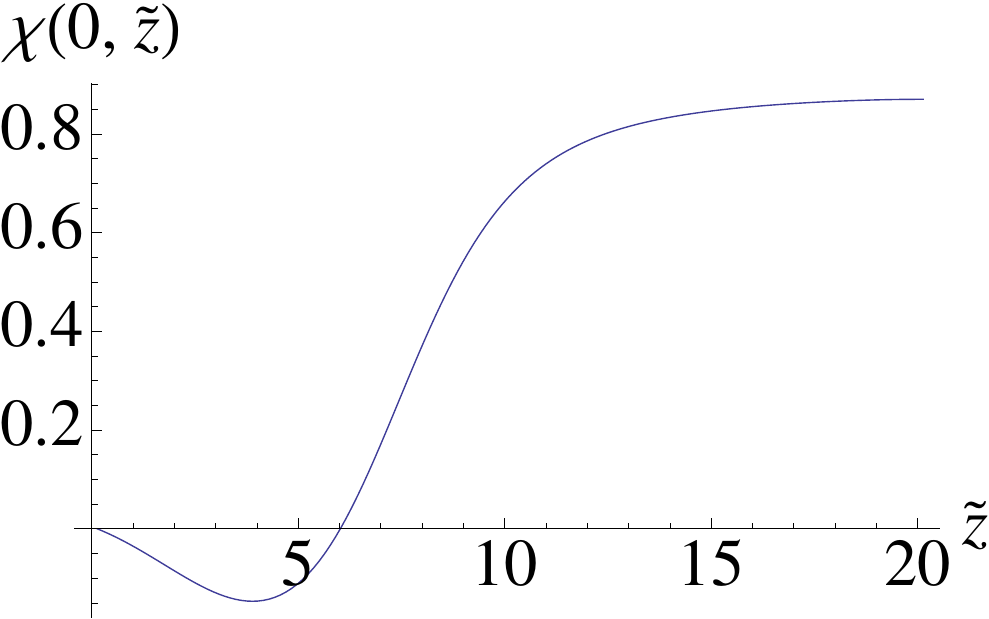}
  \caption{Interpolating $\chi$ profile}
  \label{fig:sub1}
\end{subfigure}%
\begin{subfigure}{.5\textwidth}
  \centering
  \includegraphics[width=.9\linewidth]{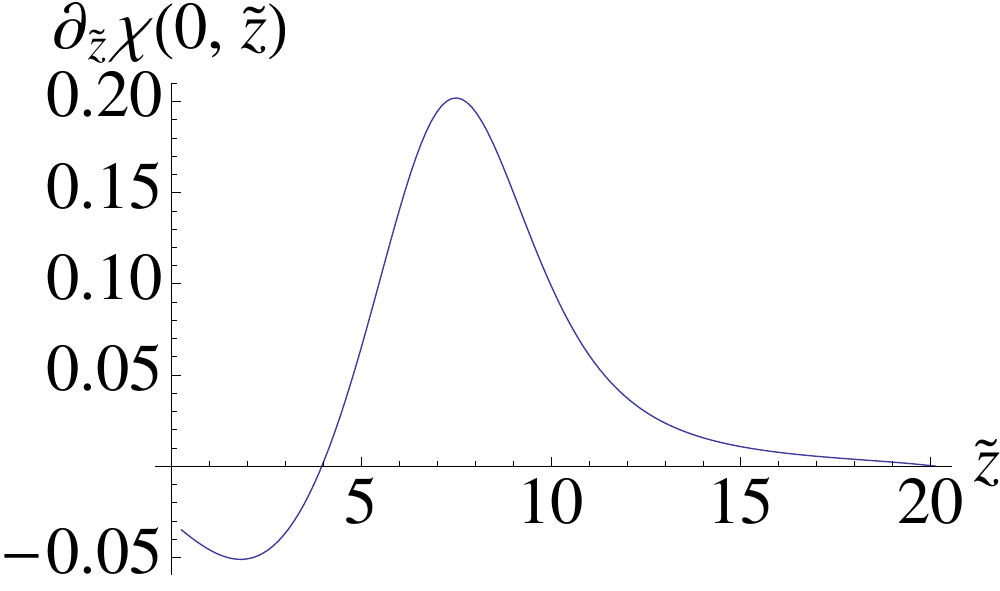}
  \caption{Derivative of $\chi$ field}
  \label{fig:sub2}
\end{subfigure}
\label{fig:test}
\centering
\begin{subfigure}{.5\textwidth}
  \centering
  \includegraphics[width=\linewidth]{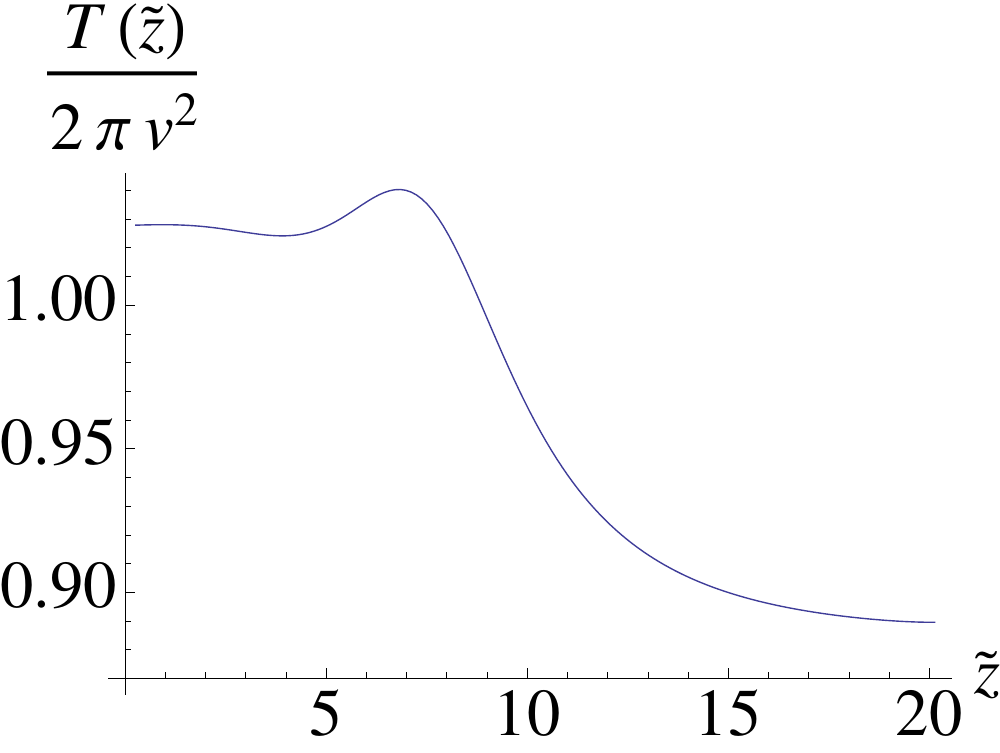}
  \caption{Energy of kink solution in $z$ direction}
  \label{fig:sub1}
\end{subfigure}%
\caption{Field profiles for the interpolating kink solution at $\rho=0$. The plots correspond to the numerical choices of $a =1, b=0.1$, $c=1.07$, $\beta = 1.1 b/(c(c-1))$, $\hat{a} = 0.53$, $n_e=1$.}
\label{fig2}
\end{figure}

The translational moduli are obvious and we shall not dwell on them further. For inclusion of the rotational moduli we allow a small $t,z$ dependence in the $\chi^i$ field in the form
\be\label{moduli1}
 \chi^i = \frac{\mu}{\sqrt{2\beta}}\chi(\rho)S^i(t,z)
 \ee
 where the moduli fields $S^i$ $(i=1,2,3)$ are constrained by
 \be
 S^iS^i = 1,
 \ee
 hence one has two moduli fields, as expected. Substituting this ansatz into (\ref{15p}) we obtain the low-energy effective action
 \be
 S= \frac{1}{2g^2}\int dtdz(\partial_k S^i)^2
\label{O3}
 \ee
 where
 \be
 \frac{1}{2g^2} = \frac{\pi}{4c\lambda\beta}I_1 \approx\frac{1.96}{\lambda},
 \ee
 with
 \be
 I_1 = \int_0^\infty \rho\chi^2(\rho)d\rho\approx3.93,
 \ee
 using the solution for $\chi(\rho)$ with large value in the string core, shown in Figure \ref{fig:sub2222}. 

Let us note that O(3) sigma model in (\ref{O3}) is a particular case of CP$(N-1)$ models (O(3) model is equivalent
to CP(1) model) which describe the internal dynamics of  non-Abelian strings in non-Abelian U$(N)$ gauge theories
\cite{AHDT}.
 

 \section{Conclusions}
 \label{con}
 
 In this paper we undertook further studies of generalized ANO strings, with non-Abelian moduli on the world sheet.
 In the simplest nonsupersymmetric model presented in Section \ref{intro}
 we found a string solution with such moduli and the tension lower than that of the ANO string.
 Then we obtained (numerically and approximately) a field configuration interpolating between the
 ANO string at one spatial infinity and the non-Abelian string at the other. This interpolating
 field exhibits a kink (although, non static due to non-degeneracy of the tensions). 
 We then calculate the kink mass and the decay rate of the ANO string into the non-Abelian string through bubble formation in the quasiclassical regime.

 \section*{Acknowledgments}

This work  is supported in part by DOE grant DE-FG02-94ER40823. G.T. would like to thank the Fine Institute for Theoretical Physics at the University of Minnesota for hospitality during the completion of this work.
The work of A.Y. was  supported 
by  FTPI, University of Minnesota, 
by RFBR Grant No. 13-02-00042a 
and by Russian State Grant for 
Scientific Schools RSGSS-657512010.2.

\end{document}